\documentclass{elsart}

\usepackage{graphicx}
\usepackage{epsfig}

\usepackage{amssymb}

\begin{document}

\begin{frontmatter}

\title{Indications of suppressed high $p_T$ hadron production in nucleus-nucleus collisions at CERN-SPS}

\author[Columbia]{David~d'Enterria}
\ead{denterria@nevis.columbia.edu}

\address[Columbia]{Nevis Laboratories, Columbia University\\ 
Irvington, NY 10533, and New York, NY 10027, USA}






\begin{abstract}
Inclusive pion production at high transverse momenta ($p_T\gtrsim$ 2 GeV/$c$) 
in nucleus-nucleus (A+A) collisions at CERN SPS 
($\sqrt{s_{\mbox{\tiny{\it{NN}}}}}\approx$ 20 GeV) is revisited and systematically 
compared to all existing proton-proton data in the same range of center-of-mass 
energies. The ratio of A+A to p+p pion cross-sections (nuclear modification factor) 
for central Pb+Pb, Pb+Au and S+Au reactions does not show a strong enhancement as 
a function of $p_T$ as previously found, but is consistent 
with scaling with the number of nucleon-nucleon ($NN$) collisions.
Neutral pion yields per $NN$ collision in head-on Pb+Pb reactions are suppressed, 
whereas peripheral yields are enhanced. These results together indicate that some 
amount of ``jet quenching'' may already be present in central heavy-ion reactions 
at $\sqrt{s_{\mbox{\tiny{\it{NN}}}}}\approx$ 20 GeV.
\end{abstract}

\end{frontmatter}

\section{Introduction}
Lattice calculations of bulk Quantum Chromodynamics (QCD) in thermal equilibrium~\cite{latt} 
predict the transition of hadronic matter to a deconfined and chirally symmetric system 
of quarks and gluons above energy densities of the order $\epsilon_{crit}\approx$ 0.7 $\pm$ 0.3 GeV/fm$^3$.
The formation and study under laboratory conditions of this ``Quark-Gluon Plasma'' 
(QGP) phase is one of the highest priorities in high-energy nuclear physics
in the present day. Several experimental results from the CERN Super Proton Synchrotron (SPS)
relativistic heavy-ion programme collected during the 1990's in fixed-target experiments 
with center-of-mass energies $\sqrt{s_{\mbox{\tiny{\it{NN}}}}}\approx$ 20 GeV have been 
interpreted, not without controversy, in terms of QGP formation~\cite{cern_press_release}. 
Indeed, although several observations in central Pb+Pb collisions~\cite{na50}
are consistent with expected QGP signals, e.g. the ``anomalous'' suppression 
of charmonium states due to Debye screening of the color potential in the plasma~\cite{satz}
(see, however, also~\cite{capella}), 
other expected signatures such as ``jet quenching'' due to parton energy loss in the dense 
deconfined medium~\cite{jet_quench} seem to be significatively absent from the data. 
In fact, high $p_T$ pion production in central A+A at CERN-SPS was found
not to be suppressed but {\it enhanced} compared to production in free space~\cite{wa98,wang_sps,wang_syst}. 
Such a ``Cronin enhancement'', observed earlier in p+A~\cite{cronin,antrea,straub} and 
$\alpha+\alpha$~\cite{ISR_pi0} collisions, is usually interpreted in terms of 
multiple initial-state parton scatterings which result in a broadening of the final $p_T$ spectra~\cite{accardi}. 
In contrast, high-$p_T$ hadro-production in central Au+Au collisions at 
$\sqrt{s_{\mbox{\tiny{\it{NN}}}}}=$ 130 GeV~\cite{phenix_hipt_130,star_hipt_130}
and 200 GeV~\cite{phenix_hipt_pi0_200,star_hipt_200,phobos_hipt_200,brahms_hipt_200} 
at the BNL Relativistic Heavy Ion Collider (RHIC) has been found to be strongly suppressed 
(by up to a factor of 4--5) compared to p+p collisions measured at the same 
$\sqrt{s_{\mbox{\tiny{\it{NN}}}}}$~\cite{phenix_pp_pi0_200,star_hipt_200}.
The observed suppression at RHIC is even more significant considering the 
fact that the ``default'' production at high $p_T$ in the ``cold'' nuclear environment 
of deuteron-nucleus reactions at collider energies is also Cronin 
enhanced~\cite{phenix_dAu,star_dAu,phobos_dAu,brahms_hipt_200}. 
These results clearly indicate that final-state effects are responsible 
for the high $p_T$ deficit observed in central Au+Au.
The difference between the {\it suppressed} RHIC and {\it enhanced} SPS 
hadro-production at large $p_T$ implies that there must exist an intermediate value of 
collision energies in nucleus-nucleus reactions at which final-state quenching
starts to dominate over initial-state $p_T$ broadening. The search for the 
onset of high $p_T$ suppression is the main motivation behind the dedicated 
$\sqrt{s_{\mbox{\tiny{\it{NN}}}}}$ = 62.4 GeV Au+Au run carried out at RHIC in April 2004.

Theoretical studies~\cite{jet_quench_review} of parton propagation in a dense medium show
that the induced parton energy loss is proportional to the initial gluon density ($dN^{g}/dy$)
in the system. In this context, the absence of suppression 
at CERN-SPS is surprising considering that experimental estimates of the 
initial energy density, based on the Bjorken prescription~\cite{bjorken} for a boost-invariant 
longitudinally expanding plasma, at a canonical thermalisation time of $\tau_0$~=~1~fm/$c$, 
are on the order of $\epsilon^{\mbox{\tiny{SPS}}}_{\mbox{\tiny{Bj}}}(\tau_0)\approx$ 3 GeV/fm$^{3}$ 
~\cite{wa98_ET,na49_ET} and 
$\epsilon^{\mbox{\tiny{RHIC}}}_{\mbox{\tiny{Bj}}}(\tau_0)\approx$ 5 GeV/fm$^{3}$ 
~\cite{phenix_ET}. 
Such energy densities, both well above the critical value of $\sim$0.7 GeV/fm$^3$, would correspond
to a ratio of parton rapidity densities ($dN/dy\propto\rho\propto\epsilon^{3/4}$ 
for an ideal gas of quarks and gluons) of $\sim$0.68 between RHIC and SPS at $\tau_0$ = 1 fm/$c$.
Since $R_{AA}^{\mbox{\tiny{RHIC}}}\approx$ 0.2, one would accordingly expect
suppression factors of the order $R_{AA}^{\mbox{\tiny{SPS}}}\approx$ 0.3 assuming 
the same Cronin $p_T$ broadening at RHIC and SPS (see discussion later). 
This is in violent contradiction with the $R_{AA}^{\mbox{\tiny{SPS}}}>>$1 usually quoted in the literature.
The usual explanations for the reported absence of suppression at SPS involve short QGP 
lifetime~\cite{wang_sps,wang_syst}, domination of multiple soft collisions (Cronin effect) 
over hard scatterings~\cite{gyul_levai_sps}, and modest amount of parton rescatterings~\cite{bass_sps}.

In this letter we explore an alternative interpretation based on a thorough reanalysis 
of the p+p $\rightarrow \pi^0$+$X$ baseline spectrum used to determine the A+A 
nuclear modification factor at $\sqrt{s}$ = 17.3 GeV in~\cite{wa98,wang_syst}.
It turns out that the absence of a concurrent proton-proton measurement at the same $\sqrt{s}$, 
and the use of inexact baseline references extrapolated from higher collision energies,
result in apparent strong Cronin enhancements which are not actually supported by the data. 
The fact that the p+p reference spectra for SPS used so far are not well under control can be 
already realized by inspecting the original work~\cite{wa98} which shows two results
difficult to reconcile at first sight: a {\it suppressed} $\pi^0$ production as given by the ratio of central 
to peripheral Pb+Pb collisions ($R_{cp}\approx$ 0.6), and a strongly {\it enhanced} central Pb+Pb 
over p+p ratio (albeit with large systematic uncertainties), $R_{AA}\approx$ 8 at the largest $p_T$. 
In this paper, the whole set of available high $p_T$ data from the SPS heavy-ion experiments: 
$\pi^0$ and $\pi^{\pm}$ at $\sqrt{s_{\mbox{\tiny{\it{NN}}}}}$ = 17.3 GeV
from Pb+Pb (WA98)~\cite{wa98} and Pb+Au (CERES/NA45)~\cite{ceres} respectively, 
and $\pi^0$ from S+Au at $\sqrt{s_{\mbox{\tiny{\it{NN}}}}}$ = 19.4 GeV (WA80)~\cite{wa80}, 
will be reexamined and carefully compared to p+p spectra constructed from a 
$\sqrt{s}$-dependent global fit of most of the available pion 
differential cross-sections in the range $\sqrt{s}\approx$ 7 -- 63 GeV~\cite{blatt}. 
Using this new reference, it will be shown that high-$p_T$ hadroproduction at 
$\sqrt{s_{\mbox{\tiny{\it{NN}}}}}\approx$ 20 GeV is not enhanced in central nucleus-nucleus 
reactions but, within errors, is consistent instead with scaling with the number of $NN$ collisions.
Furthermore, the fact that high-$p_T$ $\pi^0$ production in the the top 1\% central lead-lead
collisions appears to be suppressed by a factor of $\sim$1.6 $\pm$ 0.6 compared to the p+p reference
and that the peripheral yields are Cronin enhanced, points to some mechanism of final-state 
suppression at work in central Pb+Pb at CERN-SPS energies. The revised nuclear modification 
factors will be compared to the predictions of a pQCD-based model of parton energy loss, and 
the (partonic or hadronic) nature of the dissipative medium will be discussed.

\section{High $p_T$ pion production in p+p collisions at $\sqrt{s}\approx$ 20 GeV}

Particle production at high $p_{T}$ ($p_{T}\gtrsim$ 2 GeV/$c$) in hadronic collisions 
results from {\it incoherent} parton-parton scatterings with large $Q^2$. 
In the absence of initial and final state interactions, 
independent scattering and pQCD factorization\footnote{Incoming quarks and gluons undergoing 
hard scattering are considered ``free'' in a collinear factorized approach, i.e.
the density of partons in a nucleus with atomic number $A$ is considered to be equivalent to 
the superposition of $A$ independent nucleons, or $f_{a/A}(x,Q^2) = A\, f_{a/p}(x,Q^2)$
in terms of parton distribution functions.} imply that inclusive A+B cross-sections
for hard processes should scale simply as $A\cdot B$ times the corresponding p+p cross-sections:
$E\,d\sigma_{hard}^{AB}/d^3p=A\cdot B \cdot E\,d\sigma_{hard}^{pp}/d^3p$. 
Usually heavy-ion experiments measure invariant {\it yields} for a given centrality bin and, thus, 
the corresponding ``scaling law'' reads 
$E\,dN_{hard}^{AB}/d^3p=\langle T_{AB}(b)\rangle\cdot E\,d\sigma_{hard}^{pp}/d^3p$, where 
$T_{AB}(b)$ is the Glauber nuclear overlap function at impact parameter $b$~\cite{glauber}. 
Since the number of inelastic nucleon-nucleon collisions at $b$
is proportional to $T_{AB}$ ($N_{coll}(b) = T_{AB}(b)\cdot \sigma_{pp}^{inel}$, 
with $\sigma_{pp}^{inel}$ = 32 mb at $\sqrt{s}\approx$ 20 GeV),
one usually quantifies medium effects at high $p_{T}$ via the nuclear modification factor
\begin{equation} 
R_{AB}(p_{T})\,=\,\frac{d^2N_{AB}/dy dp_{T}}{\langle N_{coll}(b)\rangle\,\times\, d^2 N_{pp}/dy dp_{T}},
\label{eq:R_AA}
\end{equation}
which measures the deviation of A+B at impact parameter $b$ from an incoherent 
superposition of nucleon-nucleon collisions. Unfortunately, on the experimental side 
no p+p $\rightarrow \pi^0$+$X$ spectra measured at high $p_T$ close to mid-rapidity 
exists at the same collision energy\footnote{Ref.~\cite{carey} provides a 
p+p $\rightarrow \pi^0$+$X$ measurement at $p_{lab}$ = 158.9 GeV/$c$ very close 
to WA98 Pb+Pb beam energy. However, these data are inconsistent with the rest of 
the tabulated results (the authors themselves have discarded this sample from 
subsequent analysis~\cite{taylor}) and so have not been considered in this work.} 
of the SPS Pb-induced nuclear collisions
($K_{lab}$ = 158 $A$GeV corresponding to $\sqrt{s}$ = 17.3 GeV). 
On the theoretical side, the single inclusive particle spectrum at large $p_T$ in high-energy
p+p collisions can be in principle calculated within the framework of collinear factorization.
However, below $\sqrt{s}\approx$ 60 GeV as noted already in~\cite{owens77}, the 
cross-sections at $p_T<$ 5 GeV/$c$ in p+p collisions are underpredicted 
by standard pQCD calculations, and additional non-perturbative effects (e.g. intrinsic $k_T$) 
must be introduced to bring parton model analysis into agreement with data. Unfortunately, those 
effects cannot so far be introduced in a model-independent way and different pQCD 
calculations~\cite{wang_sps,gyul_levai_sps,hung_group} effectively yield different 
final pion spectra for p+p collisions around $\sqrt{s}$ = 20 GeV.

\begin{figure}[htbp]
\centerline{\psfig{figure=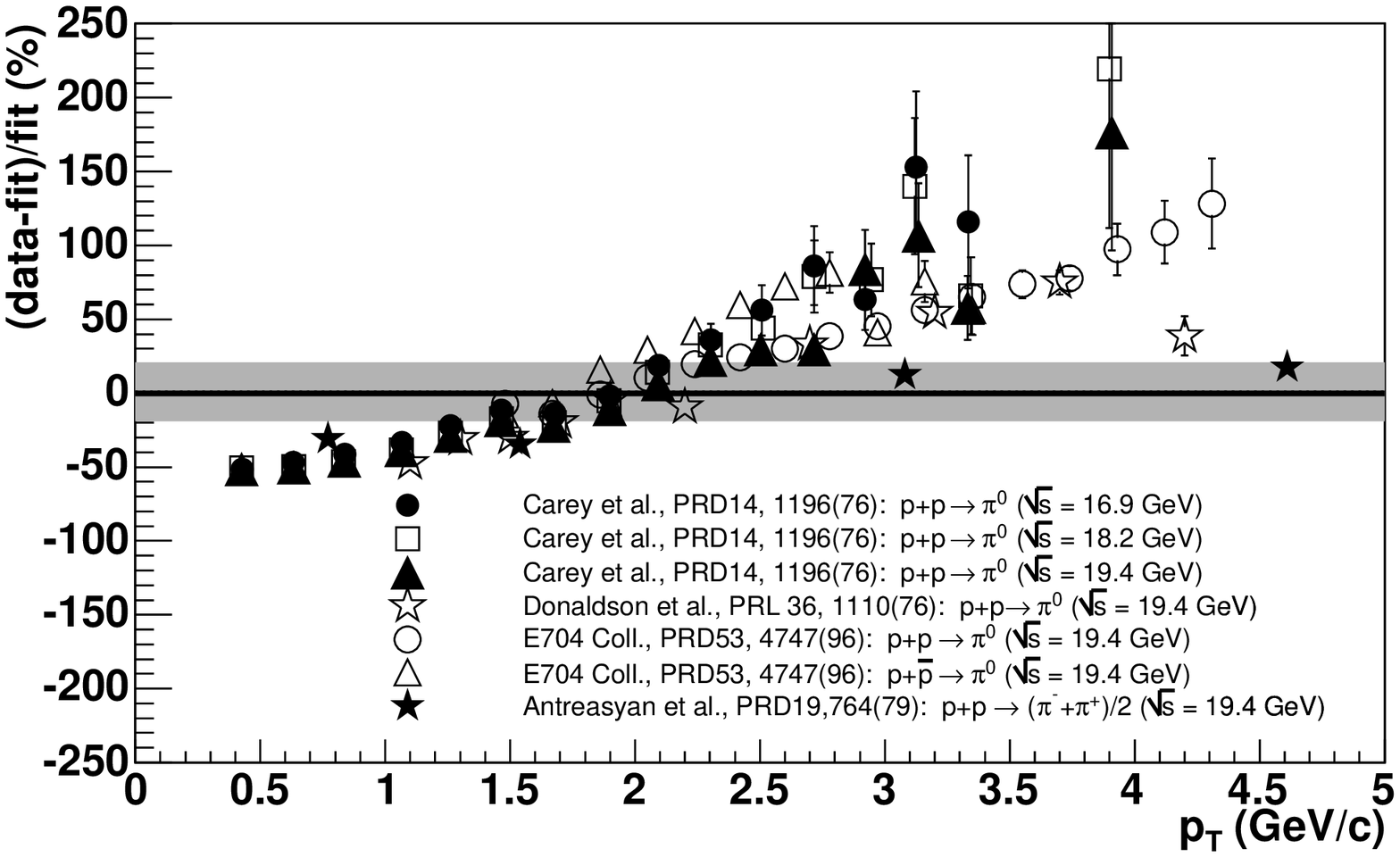,height=3.in}}
\centerline{\psfig{figure=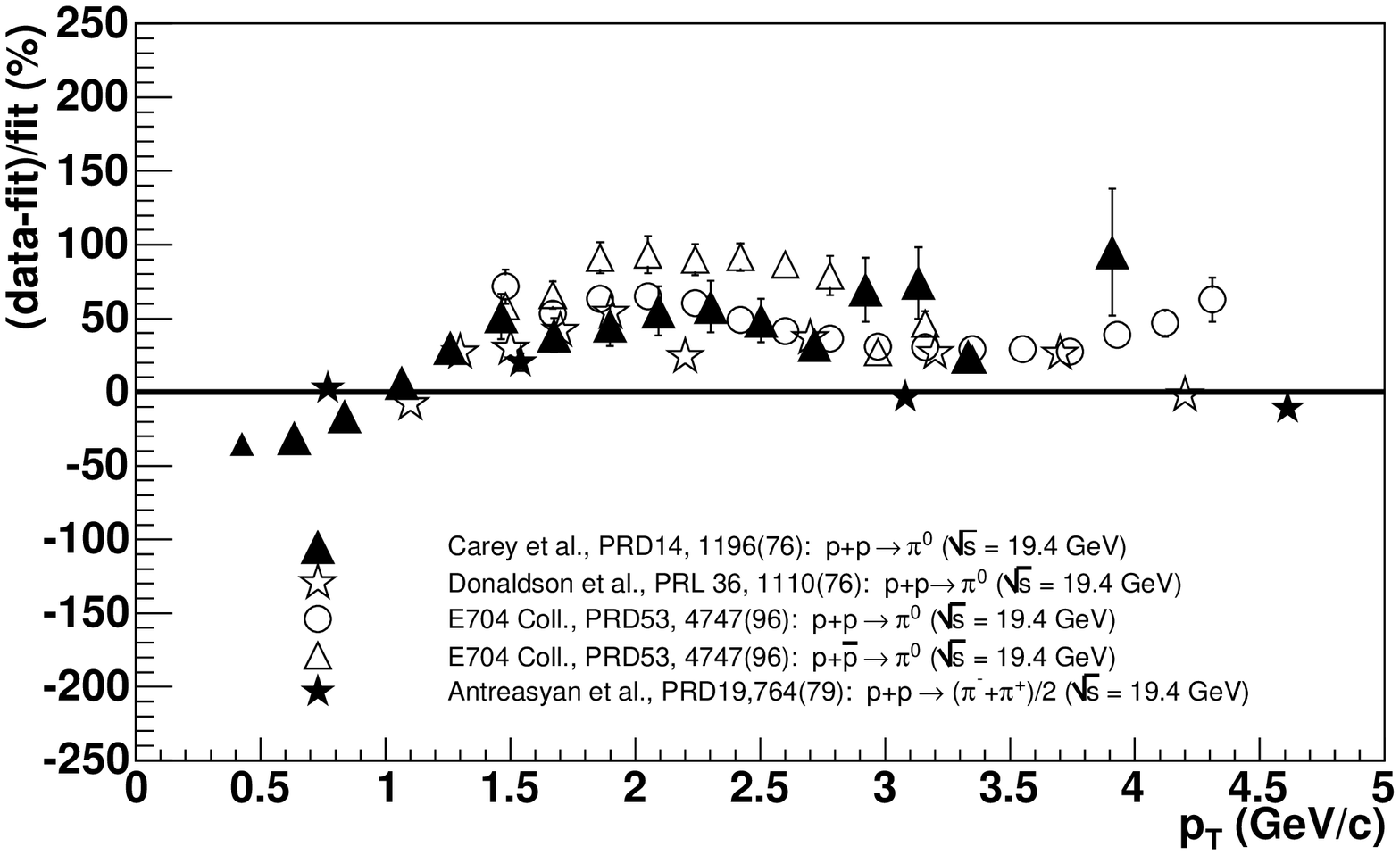,height=3.in}}
\caption{Relative differences between the single inclusive pion spectra 
measured in p+p collisions at $\sqrt{s}$ = 16.9 -- 19.4 GeV~\protect\cite{antrea,carey,donaldson,adams} 
and the p+p $\rightarrow \pi^0$+$X$ parametrizations proposed by the WA98 
collaboration~\protect\cite{wa98} (upper figure) and Wang\&Wang~\protect\cite{wang_syst}
(lower figure) at the corresponding $\sqrt{s}$. The shaded band represents the 
20\% overall uncertainty originally assigned to the WA98 parametrization.
Wang\&Wang only provides the fit parameters for 2 fixed values of $\sqrt{s}$ = 17.3, 19.4~GeV.}
\label{fig:wa98_xnwang_fits}
\end{figure}

In the absence of a concurrent experimental measurement and lacking a fully reliable theoretical
calculation, two approaches have been followed to construct a p+p reference for 
$\pi^0$ production at SPS energies. First, the WA98 collaboration~\cite{wa98} has employed 
a semi-empirical modified power-law form $A\,[p_0/(p_T+p_0)]^n$ (originally proposed by
Hagedorn~\cite{hagedorn}) tuned to reproduce the $p_T$ spectra measured at higher $\sqrt{s}$, 
plus an $x_T$ scaling prescription~\cite{beier} to account for the collision energy 
dependence of the cross section. Second, Wang\&Wang~\cite{wang_syst} have
adopted a more complex power-law ansatz for the $p_T$ spectrum which describes the 
charged pion data at $\sqrt{s}$ = 19.4 GeV~\cite{antrea}, combined with a pQCD parton 
model calculation to scale the cross-section down to $\sqrt{s}$ = 17.3 GeV.
Both parametrizations have been tuned to reproduce a subset of the existing
p+p $\rightarrow \pi$+$X$ data at $\sqrt{s}\approx$ 20 GeV, but no
true global analysis has been carried out to fully compare
the parametrizations to all the existing results in this energy regime. 
Figure~\ref{fig:wa98_xnwang_fits} shows a comparison of parametrizations~\cite{wa98} 
and~\cite{wang_syst} to the whole set of experimental results on high $p_T$ pion 
production at $\theta_{cm}\approx$ 90$^0$ in the $\sqrt{s}$ = 16.9 -- 19.4 GeV range. 
The first thing worth to notice is the relatively large disparity among the experimental data
obtained at the same energy ($\sqrt{s}$ = 19.4 GeV), especially at the largest $p_T$
values. This fact highlights the importance of concurrently measuring A+A and baseline 
p+p differential cross-sections at the same center-of-mass energy and with the same 
setup in present and future heavy-ion experiments. Having said that,
the WA98 reference fit~\cite{wa98} reproduces the available data only at $p_T\approx$ 2 GeV/$c$ 
whereas it systematically overpredicts the yields below this $p_T$ value and significantly 
underpredicts them above it. Wang\&Wang parametrization~\cite{wang_syst}, reproduces better 
the shape of the $p_T$ spectra but it systematically underpredicts most of the $\pi^0$ 
yields by $\sim$50\% in the $p_T\approx$ 1.5 -- 4.0 GeV/$c$ range. Both data fits, 
however, seem to follow rather closely the $(\pi^++\pi^-)/2$ data\footnote{Pion production in this 
kinematical range of parton fractional momenta in the proton, $x>0.2$, is dominated by valence 
$u$ (2 quarks) and $d$ (1 quark) fragmentation which, due to their respective quark content, 
results in the $\sigma(\pi^+)<\sigma(\pi^0)<\sigma(\pi^-)$ ordering of the pion cross-sections. 
One would, thus, indeed expect $(\pi^++\pi^-)/2$ to provide a good approximation of 
the $\pi^0$ yields in p+p collisions.} from Antreasyan {\it et al.}~\cite{antrea} 
which is actually used as the basic set for constraining 
the fit parameters at $\sqrt{s}$ = 19.4 GeV in~\cite{wang_syst}. Forcing the 
parametrization~\cite{wang_syst} to fit the $\pi^\pm$ data of \cite{antrea} above 
4 GeV/$c$ (where the scarce A+A data have large statistical errors anyway)
without any other constraint from existing $\pi^0$ results in the same range of 
collision energies seems to be the cause of the limited agreement 
of this parametrization with the proton-proton data in the range 
$p_T\approx$ 1.5 -- 4.0 GeV/$c$ where the heavy-ion measurements are available. 
In the case of the parametrized $\pi^0$ reference of WA98, the disagreement data-fit
seems to come from the ansatz used to take into account the $\sqrt{s}$ dependence of 
the cross section since the original parametrization reproduces well the ISR
$\pi^0$ spectra at higher energies~\cite{wa98}.

\begin{figure}[htbp]
\centerline{\psfig{figure=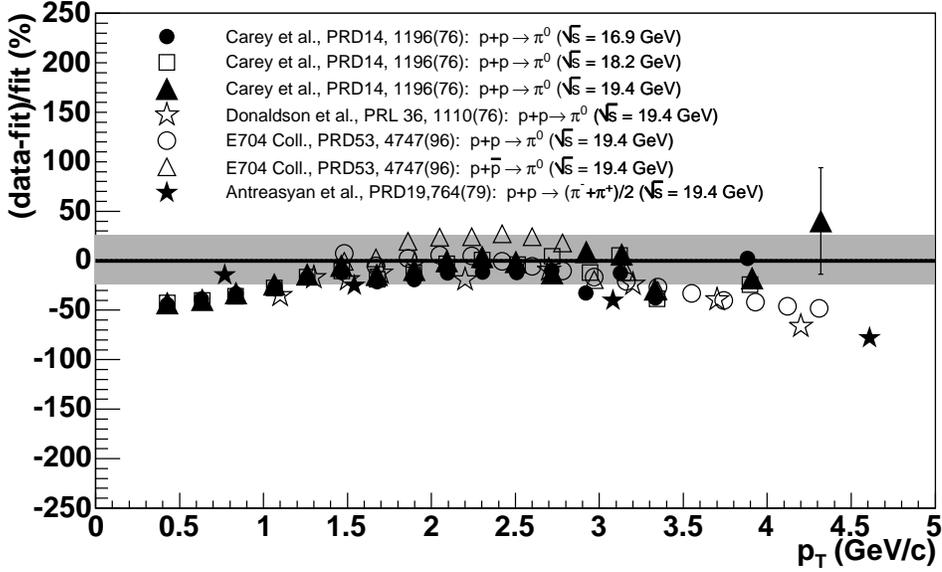,height=3.in}}
\caption{Relative differences between the single inclusive pion spectra 
measured in p+p collisions at $\sqrt{s}$ = 16.9 -- 19.4 GeV~\protect\cite{antrea,carey,donaldson,adams} 
and the p+p $\rightarrow \pi^0$+$X$ parametrization proposed in ref.~\protect\cite{blatt}.
The shaded band represents the 25\% overall uncertainty that we assign to the parametrization.}
\label{fig:fit_blatt}
\end{figure}

At variance with these two works, Blattnig {\it et. al}~\cite{blatt} derive a
parametrization of the invariant differential cross-section for inclusive $\pi^0$ 
production in proton-proton collisions based on a global analysis of most of the 
available data within $\sqrt{s}\approx$ 7 -- 63 GeV. This is a purely empirical 
11-parameter functional form tuned to provide 
a reasonable description of the full $p_T$ spectral shape, angular 
distribution and total cross-section of the measured pions in this range of center-of-mass 
energies. 
Fig.~\ref{fig:fit_blatt} shows the level of agreement of Blattnig fit to the 
available pion data in p+p collisions for beam energies in the range of CERN-SPS heavy-ion 
experiments. The agreement between data and fit is more satisfactory, both in shape and magnitude, 
than the two previous parametrizations especially within $p_T\approx$ 1.5 -- 3.5 GeV/$c$. 
It describes rather well, in particular, the most recent (and precise) data sets from Fermilab 
E704 experiment~\cite{adams} which were not actually considered in the fitting analysis 
of~\cite{blatt}. It is worth noticing, however, that at $p_T>$ 3.5 GeV/$c$ although 
the parametrization reproduces the $\pi^0$ data sets of ref.~\cite{carey}, it seems 
to be $\sim$50\% above the $\pi^0$ results of refs.~\cite{donaldson,adams} and the 
averaged $\pi^\pm$ of~\cite{antrea}. 
Nonetheless, given the relatively poor agreement between p+p experimental measurements
themselves above $p_T\approx$ 3.5 GeV/$c$, {\it and} especially given that the highest 
$p_T$ $\pi^0$ measured in A+A reactions are at $\sim$4 GeV/$c$ (with statistical
errors which are larger than the p+p reference uncertainty), we consider 
this fit to provide a much more accurate representation of the proton-proton
$\pi^0$ baseline production than the two previously used parametrizations.
Hereafter, we will thus use the reference of Blattnig {\it et. al}~\cite{blatt} 
as a reasonable estimate of the p+p$\rightarrow \pi^0$+$X$ spectra for
$\sqrt{s}$ = 17 -- 20 GeV in the range $p_T$ = 1 -- 4 GeV/$c$, -- with an 
assigned overall uncertainty of $\pm$25\% (shaded band in Fig.~\ref{fig:fit_blatt})
to cover most of the existing measurements --, as a benchmark for the study
of high $p_T$ pion production in heavy-ion reactions at SPS.

\section{Nuclear modification factors at $\sqrt{s_{\mbox{\tiny{\it{NN}}}}}\approx$ 20 GeV revisited}

\begin{figure}[htbp]
\centerline{\psfig{figure=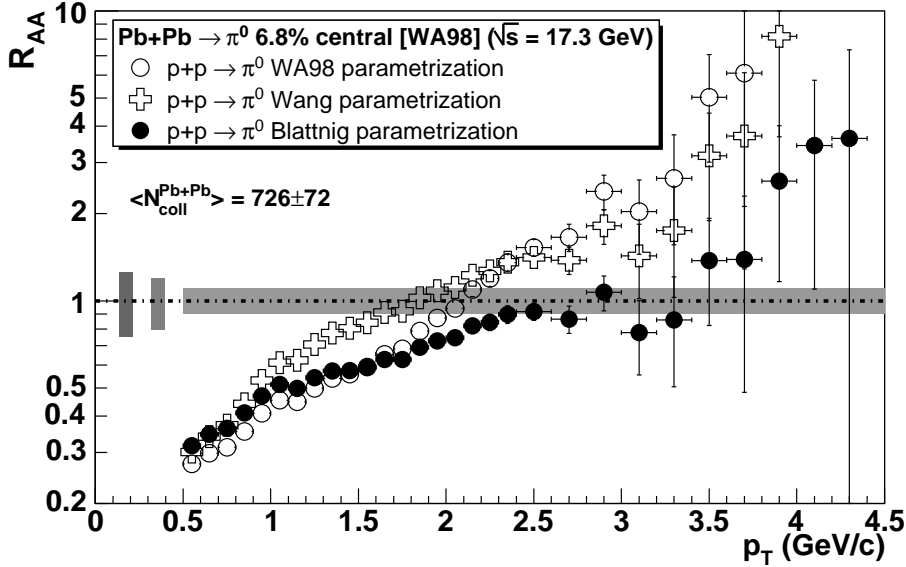,height=3.0in}}
\caption{Comparison of the nuclear modification factors, Eq. (\protect\ref{eq:R_AA}), 
for $\pi^0$ production in 0--7\% most central Pb+Pb reactions at CERN-SPS~\protect\cite{wa98} 
obtained using the three different p+p $\rightarrow\pi^0$+$X$ parametrizations~\protect\cite{wa98,wang_sps,blatt} 
discussed in the text. The shaded band around $R_{AA}$ = 1 represents the $\sim$10\% 
normalization uncertainty of the Glauber calculation of $N_{coll}$ common to all 
$R_{AA}$. The bars centered on $R_{AA}$ = 1 represent the additional fractional 
uncertainties of the Blattnig (25\%) and WA98 parametrizations (20\%) respectively.}
\label{fig:RAA_comparison}
\end{figure}

Figure~\ref{fig:RAA_comparison} shows the nuclear modification factor, Eq. (\ref{eq:R_AA}),
for 0--7\% most central Pb+Pb collisions at $\sqrt{s_{\mbox{\tiny{\it{NN}}}}}$ = 17.3 GeV,
obtained using the three p+p parametrizations discussed in the previous section. 
Particle production in the low $p_T$ region (below $p_T\approx$ 1.5 GeV/$c$)
naturally falls below the $N_{coll}$ scaling expectation ($R_{AA}$ = 1) since the 
assumption of independent point-like scattering does not hold for soft processes 
in nucleus-nucleus reactions (which instead scale with the number of participant 
nucleons in the reaction: $N_{part}\propto N_{coll}^{3/4}$~\cite{wnm}).
The original WA98 parametrization (open circles) results in a steeply rising Pb+Pb 
over p+p ratio in the whole $p_T$ range. The Cronin enhancement is apparent above 
$p_T\approx$ 2 GeV/$c$ going above $R_{AA}\approx$ 8 at the highest $p_T$'s, which
is a factor of $\sim$4 larger than the maximum pion enhancements found in p+A collisions at 
fixed-target energies ($R_{pA}\approx$ 2)~\cite{cronin,antrea,straub}. 
Wang's p+p reference (crosses) produces an overall less pronounced Cronin effect 
reaching a maximum $R_{AA}\approx$ 3 at high $p_T$. The $R_{AA}$ obtained with the Blattnig {\it et al.} fit
(closed circles) shows no indication of enhancement below $p_T\approx$ 4 GeV/$c$ and 
actually, within uncertainties, the data seem to follow the $NN$ collision scaling
($R_{AA}$ = 1) expected for hard scattering in the absence of medium effects.

Before addressing the interpretation of the observed high $p_T$ pattern in A+A 
collisions at SPS energies, it is legitimate to question the validity of the assumption 
of particle production via parton-parton scatterings for such moderate values of 
transverse momenta ($p_T\approx$ 2.0 -- 4.5 GeV/$c$). Several studies~\cite{peitzmann} 
have emphasized the relative importance of ``soft'' effects above 
$p_T\approx$ 2 GeV/$c$ in heavy-ion collisions at SPS. 
Nonetheless, there are at least four independent pieces of experimental evidence 
which seems to favor an interpretation for high $p_T$ production in nuclear 
collisions around $\sqrt{s_{\mbox{\tiny{\it{NN}}}}}$ = 20 GeV
based on hard scattering processes: 
(i) All the measured hadron $p_T$ spectra above 2 GeV/$c$~\cite{wa98,ceres,wa80} 
show a power-law tail characteristic of elementary parton-parton interactions, 
(ii) The shape and width of the near-side azimuthal correlations 
of pions with $p_T>$ 1.2 GeV/$c$ are jet-like~\cite{ceres_corr}, in agreement 
with parton fragmentation expectations, (iii) The measured ratio 
$\eta/\pi^0\approx$ 0.5 above $p_T$ = 1.5 GeV/$c$~\cite{wa80_eta} is 
consistent with standard parton fragmentation functions, and (iv) The observed 
direct photon yield at $p_T>$ 1.5 GeV/$c$~\cite{wa98_photons} 
is consistent with perturbative production cross-sections.
 
\begin{figure}[htbp]
\centerline{\psfig{figure=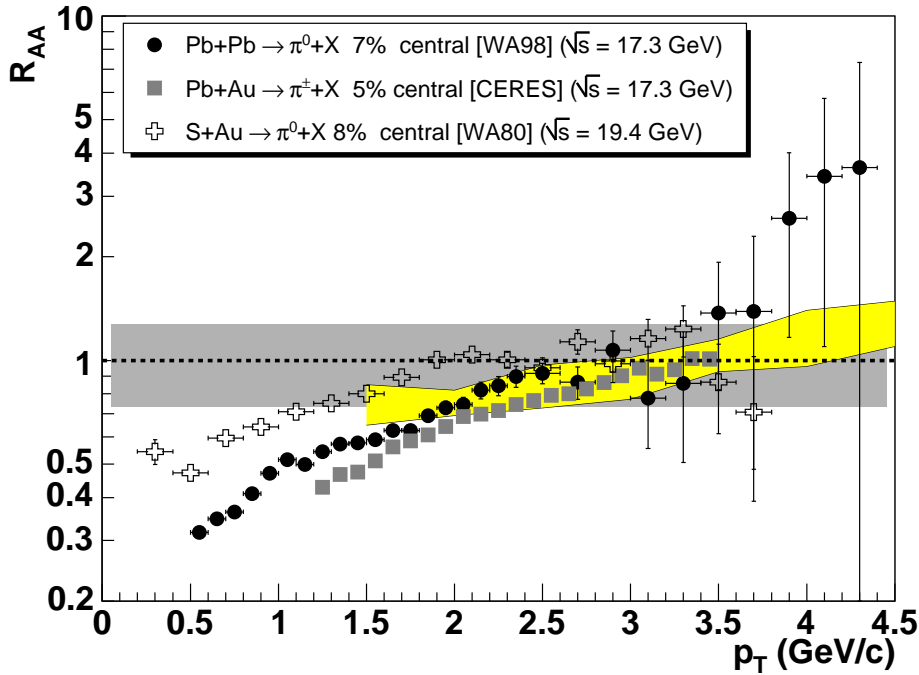,height=3.5in}}
\caption{Nuclear modification factors for pion production at CERN-SPS in 
central Pb+Pb~\protect\cite{wa98}, Pb+Au~\protect\cite{ceres}, 
and S+Au~\protect\cite{wa80} reactions, obtained using the p+p parametrization 
of ref.~\protect\cite{blatt}. The shaded band around $R_{AA}$ = 1 represents the 
overall fractional uncertainty (including in quadrature the 25\% uncertainty of
the p+p reference and the 10\% error of the Glauber calculation of $N_{coll}$: 
$\langle N_{coll}\rangle$ = 726 $\pm$ 72, 774 $\pm$ 77, and 174 $\pm$ 20 for 
Pb+Pb, Pb+Au and S+Au resp.). CERES data have an additional overall uncertainty
of $\pm$15\% not shown in the plot~\protect\cite{ceres}.
The ``curved'' band is a theoretical calculation from Vitev and 
Gyulassy~\protect\cite{vitev_gyulassy} including 
standard nuclear effects (Cronin and shadowing) and final-state parton energy 
loss in a system with initial gluon densities $dN^g/dy=$ 400 -- 600.}
\label{fig:RAA_SPS_alldata}
\end{figure}

\begin{figure}[htbp]
\centerline{\psfig{figure=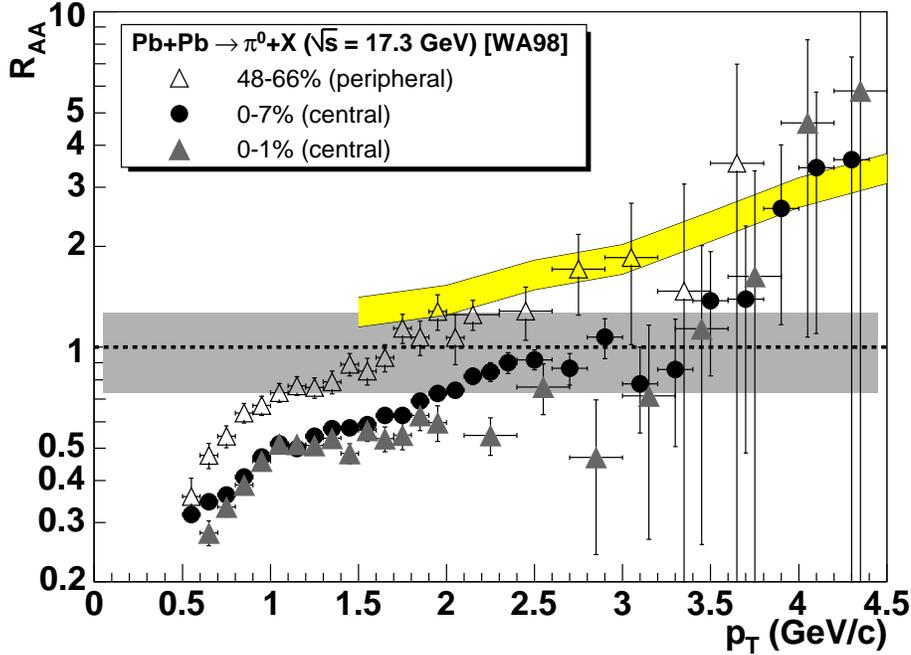,height=3.5in}}
\caption{Nuclear modification factor for $\pi^0$ production in
peripheral (48--66\%, $\langle N_{coll}\rangle$ = 78 $\pm$ 12, open triangles), 
central (0--7\%, $\langle N_{coll}\rangle$ = 726 $\pm$ 72, circles) and 
most central (0--1\%, $\langle N_{coll}\rangle$ = 807 $\pm$ 81, closed triangles) 
Pb+Pb reactions at $\sqrt{s_{\mbox{\tiny{\it{NN}}}}}$ = 17.3 GeV~\protect\cite{wa98}
obtained using the p+p baseline spectrum of ref.~\protect\cite{blatt}. 
The shaded band centered on $R_{AA}$ = 1 represents 
the overall fractional uncertainty from the p+p reference and 
Glauber calculation of $N_{coll}$. 
The curved band is a pQCD-based theoretical calculation from 
Vitev and Gyulassy~\protect\cite{vitev_gyulassy} of pion production in
central Pb+Pb at 17.3 GeV including standard nuclear effects 
(Cronin enhancement and shadowing) but {\it no} final-state parton 
energy loss.}
\label{fig:RAA_WA98_cent}
\end{figure}

Using Blattnig's parametrization~\cite{blatt} as proton-proton reference,
and the associated mean number of collisions $\langle N_{coll}\rangle$ for each centrality,
we present in Fig.~\ref{fig:RAA_SPS_alldata} the nuclear modification factors for high $p_T$ 
pion produced near midrapidity in the three A+A reactions studied at SPS 
(Table~\ref{tab:SPS_pi_data}) at comparable centrality bins. For the three systems, pion production 
in the range $p_T\approx$ 2 -- 4 GeV/$c$ is consistent with $R_{AA}$~=~1. 
Although with large experimental uncertainties, there is an indication of enhancement 
($R_{AA}>$ 1) at the highest $p_T$ values for the Pb+Pb reaction. 
The magnitude and $p_T$ dependence of the nuclear modification factors are compared
in the same plot to a theoretical calculation by Vitev and Gyulassy~\cite{vitev_gyulassy} 
(yellow band) which includes ``standard'' nuclear effects like Cronin broadening and 
(anti)shadowing, plus final-state partonic energy loss in an expanding system 
with initial gluon rapidity densities $dN^g/dy=$ 400 -- 600.
The influence of nuclear modifications (``shadowing'') of the parton distributions functions (PDFs)
in this kinematical range is small. For $\sqrt{s_{\mbox{\tiny{\it{NN}}}}}\approx$ 20 and 
$p_T\approx$ 2 -- 5 GeV/$c$ at mid-rapidity, the colliding partons have fractional momenta of the order 
$x_{Bj}=2\,P/\sqrt{s_{\mbox{\tiny{\it{NN}}}}}= 2\,z\,p_T/\sqrt{s_{\mbox{\tiny{\it{NN}}}}}\approx$ 
0.2 -- 0.4, using $\langle z\rangle=p_T/P\sim$ 0.8 for the average fraction of
the parton momentum $P$ carried by the outgoing (leading) $\pi^0$. Such an intermediate $x$ 
region is mainly dominated by valence $u,d$ quarks which are barely modified in the nucleus.
The EKS98 parametrization~\cite{eks98} of nuclear PDFs 
used by Vitev\&Gyulassy predicts a very modest $\sim$5\% antishadowing effect in this 
$p_T$ range. On the other hand, the Cronin effect does play a relevant role in $\pi^0$ 
production. The $p_T$ broadening is modeled in \cite{vitev_gyulassy} via multiple 
scattering in the cold nucleus so as to reproduce the magnitude, 
$p_T$ and $\sqrt{s}$ dependence of the enhancements in pion observed in p+A 
collisions between $\sqrt{s}\approx$ 20 -- 40 GeV~\cite{cronin,antrea,straub}. 
For SPS energies the expected effect is $R_{pA}\approx$ 1.4 at $p_T$ = 2 GeV/$c$, 
steadily increasing up to $R_{pA}\approx$ 3 at $p_T$ = 4 GeV/$c$ 
(see yellow band in Fig. \ref{fig:RAA_WA98_cent} and discussion below). In contrast 
to this expectation, the central A+A reactions at SPS show $R_{AA}\approx$~1 
within $p_T$ = 2 -- 4 GeV/$c$ (Fig.~\ref{fig:RAA_SPS_alldata}), a factor of $\sim$2 
below the expected Cronin enhancement. Introducing non-Abelian energy 
loss~\cite{jet_quench_review} of the hard scattered partons in a dense expanding 
system with initial gluon densities $dN^g/dy=$ 400 -- 600, provides the 
suppression needed to reproduce the nuclear modification factor observed in the 
A+A data in this centrality bin. Qualitatively similar conclusions have been also 
reached by closely-related pQCD-based calculations of Levai and collaborators~\cite{hung_group}.


Aside from model predictions, the presence of a final-state quenching medium in 
central A+A is confirmed by comparing high-$p_T$ $\pi^0$ production in different 
centrality bins. Fig.~\ref{fig:RAA_WA98_cent} shows the nuclear modification factor 
$R_{AA}$ versus $p_T$ for peripheral (48\%--66\% of $\sigma_{PbPb}$, open triangles), 
central (0--7\% of $\sigma_{PbPb}$, circles), and 0--1\% most central (closed triangles) 
Pb+Pb collisions measured by WA98. Pions produced in peripheral collisions above 
$p_T\approx$ 1.5 GeV/$c$ are indeed enhanced compared to ``collision scaling'', in agreement with
the phenomenological parametrization of the Cronin effect (yellow band) implemented in~\cite{vitev_gyulassy}. 
However, as aforementioned, 0--7\% most central reactions are consistent with $R_{AA}$ being unity up to 
$p_T\approx$ 3.5 GeV/$c$, and the top 1\% most central Pb+Pb reactions are actually 
found to be suppressed, $R_{AA}\approx$ 0.6 $\pm$ 0.15 (stat) $\pm$ 0.15 (syst),
in this $p_T$ range\footnote{Note that impact parameter fluctuations for this very 
narrow 0--1\% centrality bin could result in uncertainties in $R_{AA}$ larger than the 
quoted $\pm$25\% systematic error coming from the p+p reference and from the 
Glauber MC determination of the average $N_{coll}$.}. 
These results clearly indicate, regardless of the p+p $\rightarrow \pi^0$+$X$ reference used 
in the denominator of $R_{AA}$, that hard hadron production in head-on Pb+Pb collisions 
at CERN-SPS is actually quenched by a factor of $\sim$2 compared to peripheral collisions. 
It is worth to note that this was actually an observation already reported in the original 
WA98 work~\cite{wa98} which, however, remained somehow eclipsed by the (conflicting) 
enhanced values of $R_{AA}$ for all centralities quoted in the same papers.

\begin{table}[htb]
\begin{center}
\begin{tabular}{l|c|c|c|c|c|c}
\hline\hline
\hspace{1mm} 
 System [ref]& $\sqrt{s_{\mbox{\tiny{\it{NN}}}}}$ (GeV) &  $y_{cm}$ & $y_{\pi^{0}}^{exp}$ & $\epsilon_{\mbox{\tiny{Bj}}}$ 
(GeV/fm$^3$)\\\hline
 Pb+Pb $\rightarrow \pi^0$+$X$~\cite{wa98}   & 17.3 & 2.9 & 2.3$<y<$3.0  &  3.0 \cite{wa98_ET}\\
 Pb+Au $\rightarrow \pi^\pm$+$X$~\cite{ceres}& 17.3 & 2.9 & 2.1$<y<$2.6  &  3.0 \cite{na49_ET}\\
 S+Au  $\rightarrow \pi^0$+$X$~\cite{wa80}   & 19.4 & 3.0 & 2.1$\leq y\leq$2.9 & 2.0 \cite{wa80_ET}\\
\hline\hline
\end{tabular}
\label{tab:SPS_pi_data}
\end{center}
\caption{Measurements of single inclusive pion production at high $p_T$ in heavy-ion reactions 
at CERN-SPS. For each reaction we quote the center-of-mass energy and rapidity, the experimental 
rapidity coverage, and the estimated Bjorken energy density attained in the most central (0--2\%) 
collisions.}
\end{table}

\section{Discussion}

\begin{figure}[htbp]
\centerline{\psfig{figure=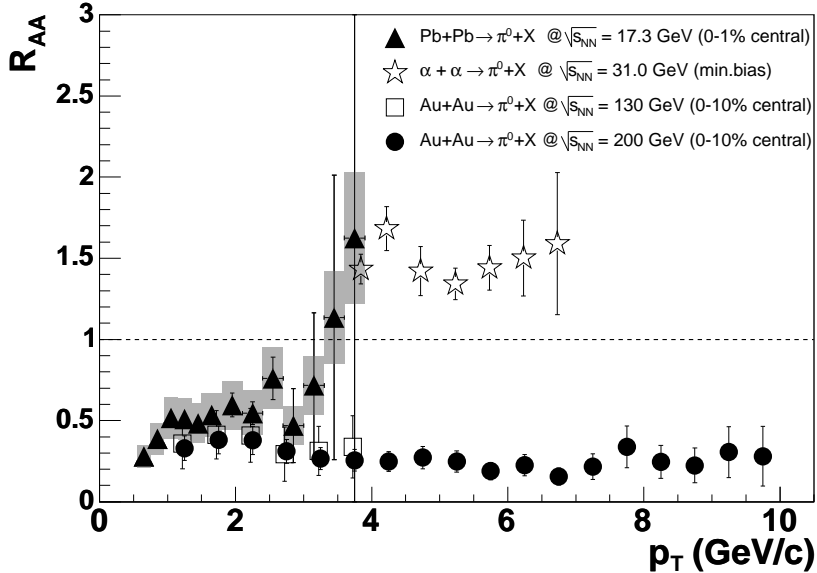,height=3.in}}
\caption{
Nuclear modification factor, $R_{AA}(p_T)$, for $\pi^0$ production 
in ion-ion reactions at CERN-SPS~\protect\cite{wa98} (triangles), 
CERN-ISR~\protect\cite{ISR_pi0} (stars), 
and BNL-RHIC~\protect\cite{phenix_hipt_130,phenix_hipt_pi0_200} 
(squares and circles). The boxes around the CERN-SPS data points 
represent the normalization uncertainty from the p+p reference and
Glauber calculation of $N_{coll}$.}
\label{fig:RAA_RHIC_ISR_SPS}
\end{figure}

Figure~\ref{fig:RAA_RHIC_ISR_SPS} shows the nuclear modification
factors for $\pi^0$ production in nucleus-nucleus reactions
at four different center-of-mass energies and different centralities.
Single inclusive pion spectra above $p_T\approx$ 2 GeV/$c$ produced at 
midrapidity in heavy-ion reactions at SPS and RHIC are suppressed 
by as much as a factor of $\sim$1.6 $\pm$ 0.6 (in top 0--1\% central Pb+Pb at SPS) 
and of $\sim$5 $\pm$ 1 (in 0--10\% most central Au+Au at RHIC) respectively, 
compared to proton-proton reactions scaled by the corresponding number 
of $NN$ collisions. In contrast, high $p_T$ $\pi^0$ production in minimum-bias 
light-ion ($\alpha+\alpha$) reactions at ISR energies is enhanced 
($R_{AA}\approx$ 1.5) with respect to this scaling. 
Such observations are consistent with the expectations of final-state 
energy loss of the hard scattered partons in dense strongly interacting matter 
produced at midrapidity in central reactions with heavy nuclei. Determining 
whether the quenching medium is of partonic or hadronic nature (or both) is 
the ultimate goal behind the study of high $p_T$ hadroproduction in high-energy 
heavy-ion collisions. The SPS and RHIC measurements of the total transverse energy 
at central rapidities yield values of the Bjorken energy density at $\tau_0$~=~1~fm/$c$, 
$\epsilon^{\mbox{\tiny{SPS}}}_{\mbox{\tiny{Bj}}}(\tau_0)\approx$ 3 GeV/fm$^{3}$
and $\epsilon^{\mbox{\tiny{RHIC}}}_{\mbox{\tiny{Bj}}}(\tau_0)\approx$ 5 GeV/fm$^{3}$, 
in the top 2\% central Pb+Pb,Au~\cite{wa98_ET,na49_ET} and top 5\% central 
Au+Au~\cite{phenix_ET} reactions resp., which are well above any possible scenario 
involving hadronic degrees of freedom. The equivalent parton densities are 
$\rho^{\mbox{\tiny{SPS}}}(\tau_0)\approx$ 4.7 fm$^{-3}$ 
and $\rho^{\mbox{\tiny{RHIC}}}(\tau_0)\approx$ 6.9 fm$^{-3}$ using the 
thermodynamical relation $\rho \approx 2.1\,\epsilon^{3/4}$ 
($\epsilon$ in GeV/fm$^{3}$) for a gas of partons as given from lattice QCD 
thermodynamics\footnote{Mind that the use of the ideal gas EOS from lattice calculations 
with zero baryochemical potential may be justified at central rapidities for RHIC 
highest energies (where $\mu_{B} << T_{c}$) but is less evident for SPS where 
$\mu_{B}\gtrsim T_{c}$, and should be taken {\it cum grano salis} in this latter case.}~\cite{latt}. 
The corresponding parton densities per unit rapidity, 
$dN/dy = \rho \cdot \tau_0 \cdot A_\perp$, where $A_\perp\approx$ 150 fm$^2$ is the 
transverse area in a head-on A+A collision with A $\approx$ 200, are
$(dN/dy)^{\mbox{\tiny{SPS}}}(\tau_0)\approx$ 700 and 
$(dN/dy)^{\mbox{\tiny{RHIC}}}(\tau_0)\approx$ 1100 respectively. Both values
are consistent with the initial $dN^{g}/dy$ gluon densities obtained from the respective 
``tomographic''parton energy loss studies~\cite{vitev_gyulassy} 
described before\footnote{The values $(dN^{g}/dy)^{\mbox{\tiny{SPS}}}(\tau_0^{_{'}})\approx$ 600 
and $(dN^{g}/dy)^{\mbox{\tiny{RHIC}}}(\tau_0^{_{'}})\approx$ 1100~\cite{vitev} 
are actually obtained for $\tau_0^{_{'}\mbox{\tiny{SPS}}}$ = 0.8 fm/$c$ and 
$\tau_0^{_{'}\mbox{\tiny{RHIC}}}$ = 0.6 fm/$c$ respectively, for a system with 
transverse area $A_\perp\approx$ 115 fm$^2$. However, when scaled to 
the same area and initial proper time (taking into account the decrease in 
energy density $\epsilon\propto\tau^{-4/3}$ in a 1+1D Bjorken expansion) 
one gets coincident results between $dN^{g}/dy(\tau_0^{_{'}})$ (from tomographic
studies) and $dN/dy(\tau_0$ = 1 fm/$c$) (from Bjorken energy densities).}.

Since jet quenching models predict that the induced parton energy loss 
is proportional to the initial parton densities, one could use the measured suppression at RHIC
together with the relative values of $\epsilon_{Bj}$, or $dN/dy(\tau_0)$, 
between RHIC and SPS to determine the expected value of quenching factor $R_{AA}$ at SPS.
Indeed, the fact that high-$p_T$ hadron spectra in proton-proton collisions are well 
reproduced by a simple power-law $1/p_T^n$ (with $n\approx$ 10 above $p_T\approx$ 2 GeV/$c$ 
at SPS, and $n\approx$ 8 above $p_T\approx$ 4 GeV/$c$ at RHIC), and that $R_{AA}$ in central 
A+A reactions is approximately constant at high-$p_T$ both at SPS and RHIC, indicates that 
the basic $p_T$ dependence of the yields is not changed by the quenching medium 
(i.e. the power law $n$ exponent remains the same). We can thus relate the suppression 
in A+A compared to p+p as due to a corresponding (energy loss) shift $\Delta p_T$ in the 
single inclusive spectrum $dN/dp_T$. The ratio of A+A over p+p invariant $dN/p_Tdp_T$ spectra, 
the nuclear modification factor, is then $R_{AA}=(1+\Delta p_T/p_T)^{-(n-1)}$, from which 
the corresponding fractional energy loss can be derived: 
$\Delta p_T/p_T = R_{AA}^{-1/(n-1)}-1$. Thus, from the measured $R_{AA}^{\mbox{\tiny{RHIC}}}\approx$ 0.2
one gets $(\Delta p_T/p_T)_{\mbox{\tiny{RHIC}}}\approx$ 0.25, and since 
$(dN/dy)^{\mbox{\tiny{SPS}}}\approx 0.68\,(dN/dy)^{\mbox{\tiny{RHIC}}}$ implies
$(\Delta p_T)_{\mbox{\tiny{SPS}}}\approx 0.68\,(\Delta p_T)_{\mbox{\tiny{RHIC}}}$, 
one would expect $R_{AA}^{\mbox{\tiny{SPS}}}\approx$ 0.4. The minimum value of 
$R_{AA}^{\mbox{\tiny{SPS}}}$ measured is, however, 
$R_{AA}^{\mbox{\tiny{SPS}}}\approx$ 0.6 for the 1\% most central Pb+Pb reactions 
(Fig. \ref{fig:RAA_WA98_cent}) which is $\sim$70\% larger than this 
simple estimate. The reason for this apparent inconsistency is the implicit 
assumption made above that the counteracting effect of the (Cronin) $p_T$ broadening 
is the same at SPS and RHIC energies. Since the high-$p_T$ spectra at SPS (with power 
law exponent $n\approx$ 10) are much steeper than at RHIC ($n\approx$ 8) the effect of 
the initial-state multiple parton scatterings leads to a much larger $p_T$ 
enhancement at $\sqrt{s_{\mbox{\tiny{\it{NN}}}}}\approx$ 20 GeV than at 
$\sqrt{s_{\mbox{\tiny{\it{NN}}}}}$ = 200 GeV. This fact, as pointed out by
Gyulassy and Levai in~\cite{gyul_levai_sps}, explains partially why the 
observed suppression at SPS is apparently much lower than at RHIC even though 
the estimated energy densities are only a factor of $\sim$2 larger at RHIC.

At both center-of-mass energies, however, it is conceivable that not all of the 
high-$p_T$ hadron suppression in central heavy-ion reactions is due to the 
attenuating effects of a {\it partonic} medium alone. 
Indeed, since the produced medium undergoes a longitudinal expansion, its initial 
energy density  will decrease with time as $\epsilon\propto(\tau_0/\tau)^{\alpha}$ 
where $\alpha$ = 1, 4/3 for free streaming and 1-D Bjorken expansion\footnote{The 
assumption of pure Bjorken (longitudinal boost invariant) expansion is an idealistic 
approximation that can be only applied in heavy-ion collisions at best in a narrow
rapidity window around $y$ = 0.} respectively~\cite{wang_sps} in the schematic
time-scale evolution outlined here. 
Therefore, even starting at $\tau_0$~=~1~fm/$c$ with energy densities well above 
$\epsilon_{crit}\approx$ 0.7 GeV/fm$^{3}$,
the bulk partonic system will drop below $\epsilon_{crit}$ and hadronize 
into a hadron gas phase at $\tau_{crit}^{\mbox{\tiny{\it{SPS}}}}\approx$ 4~fm/$c$
and $\tau_{crit}^{\mbox{\tiny{\it{RHIC}}}}\approx$ 7~fm/$c$ respectively. 
The subsequent hadron system remains strongly self-interacting until its density 
is too low for further rescatterings to take place. 
At both center-of-mass energies, the total lifetime of the strongly 
interacting system ($\tau_{fo}\approx$ 15~fm/$c$~\cite{freezeout_sps,freezeout_rhic})
is comparable to the time it takes a hard scattered parton 
with typical momentum $P$ = 4 (8) GeV/$c$ to hadronize into a fully formed 
meson~\cite{dokshitzer}: $\tau_h\approx P\cdot R_h^2\approx$ 12 (25) fm/$c$
(for a pion of radius $R_h\sim$ 0.8 fm~\cite{pion_radius}). 
In this context, the produced high-energy parton travels (and loses energy) first 
through a dense partonic system during $\tau<\tau_{crit}$ and then, for a time 
$\tau_{crit}<\tau<\tau_{fo}$, through a hadronic environment. 
In this second hadronic stage, inelastic scattering of the ``pre-hadron'' object with
comover soft hadrons of the type described in~\cite{capella,gallmeister}
can also partially account for the suppression of the final high-$p_T$ inclusive 
spectra. In any case, it is reasonable to admit that the energy loss will be larger 
in the denser (partonic) phase than in the more rarefied hadronic one. 



\section{Conclusions}
We have reexamined high-$p_T$ ($p_T\gtrsim$ 2 GeV/$c$) inclusive pion 
production in nucleus-nucleus reactions from CERN-SPS fixed-target experiments 
at center of mass energies around $\sqrt{s_{\mbox{\tiny{\it{NN}}}}}$ = 20 GeV, 
and systematically compared them to the available proton-proton 
data in the same range of collision energies per nucleon-nucleon pair. 
In contrast to what has been usually considered so far, we conclude that there is 
no indication of a strong $p_T$ broadening (Cronin enhancement) in the high-$p_T$ 
yields measured in central Pb+Pb, Pb+Au and S+Au reactions. 
Instead, the data appear to be consistent within errors with the perturbative 
expectations of scaling with the number of nucleon-nucleon collisions. The peripheral 
yields, however, are still found to be Cronin enhanced. These facts, together with 
the complementary observation that high-$p_T$ pion yields in head-on (0--1\% most central) 
Pb+Pb reactions are suppressed by a factor of $\sim$1.6 $\pm$ 0.6 compared to p+p collisions, 
is consistent with a moderate amount of final-state quenching of the hard scattered 
partons traversing the dense system produced in the course of the most central 
heavy-ion reactions at SPS. Theoretical calculations of parton energy loss in 
an expanding deconfined medium require initial gluon rapidity densities of the 
order $dN^g/dy\approx$ 600 to reproduce the observed yields, consistent with 
estimations of the Bjorken energy densities attained in the reactions. 
Additionally, we have provided arguments based on hadronization time estimates that support 
the idea that the hard scattered partons must lose their energy in the dense partonic 
{\it and} hadronic phases of the reaction. The 2004 $\sqrt{s_{\mbox{\tiny{\it{NN}}}}}$ = 62.4 GeV 
Au+Au run at RHIC will undoubtedly help to clarify the ``excitation function'' evolution
of the high-$p_T$ hadron suppression observed in high energy nucleus-nucleus reactions.
The discussion presented here highlights the importance of a concurrent and precise 
measurement of the high-$p_T$ production yields in baseline p+p collisions at the 
same center-of-mass energies as the nucleus-nucleus data.

{\it Note added: Preliminary RHIC results from Au+Au reactions at 
$\sqrt{s_{\mbox{\tiny{\it{NN}}}}}$ = 62.4 GeV, obtained after submission of this letter, 
indicate that high $p_T$ hadron production per $NN$ collision is also significantly 
reduced (by up to a factor of $\sim$3) in central Au+Au compared to p+p collisions 
at these energies. These results are consistent with the indications of moderate
high $p_T$ suppression in heavy-ion reactions at lower SPS energies, discussed here.}

\section{Acknowledgments}

The author would like to thank K.~Reygers, T.~Peitzmann and J.~Bielcikova for providing the
WA98, WA80 and CERES data respectively and for helpful discussions, as well as to 
C.~Blume and P.~Stankus for valuable comments. I am indebted to P.~Stankus 
for a careful reading of a previous version of this paper, as well as to I.~Vitev 
for making available his parton energy loss calculations and for useful 
remarks.





\end{document}